\documentclass[12pt]{article}
\usepackage{epsfig}
\usepackage{amsmath}
\usepackage{hhline}
\usepackage{amssymb}
\usepackage{times}
\usepackage{cite}
\newlength{\dinwidth}
\newlength{\dinmargin}
\usepackage{graphicx} 
\usepackage{lscape} 
\usepackage{fancybox} 
\usepackage{verbatim} 
\usepackage{booktabs}
\begin{document}

%%%%%%%%%%%%%%%%
\newcommand{\pom}{{I\!\!P}}
\newcommand{\reg}{{I\!\!R}}
\newcommand{\slowpi}{\pi_{\mathit{slow}}}
\newcommand{\fiidiii}{F_2^{D(3)}}
\newcommand{\fiidiiiarg}{\fiidiii\,(\beta,\,Q^2,\,x)}
\newcommand{\n}{1.19\pm 0.06 (stat.) \pm0.07 (syst.)}
\newcommand{\nz}{1.30\pm 0.08 (stat.)^{+0.08}_{-0.14} (syst.)}
\newcommand{\fiidiiiful}{F_2^{D(4)}\,(\beta,\,Q^2,\,x,\,t)}
\newcommand{\fiipom}{\tilde F_2^D}
\newcommand{\ALPHA}{1.10\pm0.03 (stat.) \pm0.04 (syst.)}
\newcommand{\ALPHAZ}{1.15\pm0.04 (stat.)^{+0.04}_{-0.07} (syst.)}
\newcommand{\fiipomarg}{\fiipom\,(\beta,\,Q^2)}
\newcommand{\pomflux}{f_{\pom / p}}
\newcommand{\nxpom}{1.19\pm 0.06 (stat.) \pm0.07 (syst.)}
\newcommand {\gapprox}
   {\raisebox{-0.7ex}{$\stackrel {\textstyle>}{\sim}$}}
\newcommand {\lapprox}
   {\raisebox{-0.7ex}{$\stackrel {\textstyle<}{\sim}$}}
\def\gsim{\,\lower.25ex\hbox{$\scriptstyle\sim$}\kern-1.30ex%
\raise 0.55ex\hbox{$\scriptstyle >$}\,}
\def\lsim{\,\lower.25ex\hbox{$\scriptstyle\sim$}\kern-1.30ex%
\raise 0.55ex\hbox{$\scriptstyle <$}\,}
\newcommand{\pomfluxarg}{f_{\pom / p}\,(x_\pom)}
\newcommand{\dsf}{\mbox{$F_2^{D(3)}$}}
\newcommand{\dsfva}{\mbox{$F_2^{D(3)}(\beta,Q^2,x_{I\!\!P})$}}
\newcommand{\dsfvb}{\mbox{$F_2^{D(3)}(\beta,Q^2,x)$}}
\newcommand{\dsfpom}{$F_2^{I\!\!P}$}
\newcommand{\gap}{\stackrel{>}{\sim}}
\newcommand{\lap}{\stackrel{<}{\sim}}
\newcommand{\fem}{$F_2^{em}$}
\newcommand{\tsnmp}{$\tilde{\sigma}_{NC}(e^{\mp})$}
\newcommand{\tsnm}{$\tilde{\sigma}_{NC}(e^-)$}
\newcommand{\tsnp}{$\tilde{\sigma}_{NC}(e^+)$}
\newcommand{\st}{$\star$}
\newcommand{\sst}{$\star \star$}
\newcommand{\ssst}{$\star \star \star$}
\newcommand{\sssst}{$\star \star \star \star$}
\newcommand{\tw}{\theta_W}
\newcommand{\sw}{\sin{\theta_W}}
\newcommand{\cw}{\cos{\theta_W}}
\newcommand{\sww}{\sin^2{\theta_W}}
\newcommand{\cww}{\cos^2{\theta_W}}
\newcommand{\trm}{m_{\perp}}
\newcommand{\trp}{p_{\perp}}
\newcommand{\trmm}{m_{\perp}^2}
\newcommand{\trpp}{p_{\perp}^2}
\newcommand{\alp}{\alpha_s}

\newcommand{\alps}{\alpha_s}
\newcommand{\sqrts}{$\sqrt{s}$}
\newcommand{\LO}{$O(\alpha_s^0)$}
\newcommand{\Oa}{$O(\alpha_s)$}
\newcommand{\Oaa}{$O(\alpha_s^2)$}
\newcommand{\PT}{p_{\perp}}
\newcommand{\JPSI}{J/\psi}
\newcommand{\sh}{\hat{s}}
\newcommand{\uh}{\hat{u}}
\newcommand{\MP}{m_{J/\psi}}
\newcommand{\PO}{I\!\!P}
\newcommand{\xbj}{x}
\newcommand{\xpom}{x_{\PO}}
\newcommand{\ttbs}{\char'134}
\newcommand{\xpomlo}{3\times10^{-4}}  
\newcommand{\xpomup}{0.05}  
\newcommand{\dgr}{^\circ}
\newcommand{\pbarnt}{\,\mbox{{\rm pb$^{-1}$}}}
\newcommand{\gev}{\,\mbox{GeV}}
\newcommand{\WBoson}{\mbox{$W$}}
\newcommand{\fbarn}{\,\mbox{{\rm fb}}}
\newcommand{\fbarnt}{\,\mbox{{\rm fb$^{-1}$}}}
\newcommand{\dsdx}[1]{$ d\sigma\!/\!d #1\,$}
\newcommand{\dsd}[1]{${\rm d}\sigma\!/\!{\rm d} #1\,$}
\newcommand{\dds}{$d\sigma\!/\!d x\, d Q^2\,$}
\newcommand{\eV}{\mbox{e\hspace{-0.08em}V}}
%
% Some useful tex commands
%
\newcommand{\qsq}{\ensuremath{Q^2} }
\newcommand{\gevsq}{\ensuremath{\mathrm{GeV}^2} }
\newcommand{\et}{\ensuremath{E_t^*} }
\newcommand{\rap}{\ensuremath{\eta^*} }
\newcommand{\gp}{\ensuremath{\gamma^*}p }
\newcommand{\dsiget}{\ensuremath{{\rm d}\sigma_{ep}/{\rm d}E_t^*} }
\newcommand{\dsigrap}{\ensuremath{{\rm d}\sigma_{ep}/{\rm d}\eta^*} }

%%% Dstar stuff
\newcommand{\dstar}{\ensuremath{D^*}}
\newcommand{\dstarp}{\ensuremath{D^{*+}}}
\newcommand{\dstarm}{\ensuremath{D^{*-}}}
\newcommand{\dstarpm}{\ensuremath{D^{*\pm}}}
\newcommand{\zDs}{\ensuremath{z(\dstar )}}
\newcommand{\Wgp}{\ensuremath{W_{\gamma p}}}
\newcommand{\ptds}{\ensuremath{p_t(\dstar )}}
\newcommand{\etads}{\ensuremath{\eta(\dstar )}}
\newcommand{\ptj}{\ensuremath{p_t(\mbox{jet})}}
\newcommand{\ptjn}[1]{\ensuremath{p_t(\mbox{jet$_{#1}$})}}
\newcommand{\etaj}{\ensuremath{\eta(\mbox{jet})}}
\newcommand{\detadsj}{\ensuremath{\eta(\dstar )\, \mbox{-}\, \etaj}}

% Journal macro
\def\Journal#1#2#3#4{{#1} {\bf #2} (#3) #4}
\def\NCA{ Nuovo Cimento}
\def\NIM{ Nucl. Instrum. Methods}
\def\NIMA{{ Nucl. Instrum. Methods} {\bf A}}
\def\NPB{{ Nucl. Phys.}   {\bf B}}
\def\PLB{{ Phys. Lett.}   {\bf B}}
\def\PRL{ Phys. Rev. Lett.}
\def\PRD{{ Phys. Rev.}    {\bf D}}
\def\ZPC{{ Z. Phys.}      {\bf C}}
\def\EJC{{ Eur. Phys. J.} {\bf C}}
\def\CPC{ Comp. Phys. Commun.}

%%%%%%%%%%%%%%%% extra stuff

\newcommand{\ord}{{\cal O}}
\renewcommand{\L}{{\cal L}}
\newcommand{\rpv}{\slash\hspace{-3.1mm}{R}_{p}}
\newcommand{\brpv}{\slash\hspace{-3.8mm}{R}_{p}}

\newcommand{\GeV}{\mathrm{GeV}}
\newcommand{\TeV}{\mathrm{TeV}}
\newcommand{\pb}{\mathrm{pb}}
\newcommand{\cm}{\mathrm{cm}}
\newcommand{\m}{\mathrm{m}}
\newcommand{\hdick}{\noalign{\hrule height1.4pt}}
\newcommand{\nn}{\nonumber}
\newcommand{\beq}{\begin{equation}}
\newcommand{\eeq}{\end{equation}}
\newcommand{\bea}{\begin{eqnarray}}
\newcommand{\eea}{\end{eqnarray}}
\newcommand{\eq}[1]{eq.~(\ref{#1})}

\newcommand{\tf}[1]{\parbox{8mm}{\hfill #1}}

%%%%%%%%%%%%%%%%%%%%%%%%%%%%%%%%%%%%%%%%%%
\begin{titlepage}

\noindent
\begin{flushleft}
{\tt DESY 11-114    \hfill    ISSN 0418-9833} \\
{\tt July 2011}                  \\
\end{flushleft}

\noindent
%Date:          \\ \today      \\
%Version:       2.9 \\
%%.2...; 1st draft: 1.0, 1.1...; 2nd Draft 2.0..., Final Reading 3.0,3.1...      \\
%Editors: H.~Pirumov, V.~Radescu, A.~Sch\"oning           \\
%Referees: J.~List, R.~Placakyte          \\
%Final Reading Friday, July 1st, 8h30, spokesperson office       \\
\noindent
\vspace{2cm}

\begin{center}
  \begin{Large}{\bf
      Search for Contact Interactions \\  in {\boldmath 
      $e^\pm p$} Collisions at HERA}
    \\ \vspace*{2cm}
    {H1 Collaboration}
  \end{Large} 
\end{center}

\vspace{2cm}

\begin{abstract}
A search for physics beyond the Standard Model in neutral current deep
inelastic scattering at high negative four-momentum transfer  
squared \qsq is performed in $e^{\pm}p$ collisions at HERA. 
The differential cross section \dsd {\qsq}, measured using the
full H1 data sample corresponding to an integrated luminosity of
$446$~pb$^{-1}$, is compared to the Standard Model prediction. 
No significant deviation is observed.
Limits on various models predicting new phenomena at high \qsq are derived.
For general four-fermion $eeqq$ contact interaction models, lower limits on the compositeness scale $\Lambda$ are set in the range $3.6~\TeV$ to $7.2~\TeV$.
Leptoquarks with masses $M_{\rm LQ}$ and couplings $\lambda$ are
constrained to $M_{\rm LQ}/\lambda > 0.41-1.86~\TeV$ and limits on
squarks in $R$-parity violating supersymmetric models are
derived.
A lower limit on the %effective Planck 
gravitational scale in (4+n) dimensions of
 %$M_S > 0.9~\TeV$ 
 $M_{S} > 0.9~\TeV$ 
is established for
low-scale quantum gravity effects in models with large extra dimensions. 
For the light quark radius an upper bound of $R_q< 0.65 \cdot 10^{-18}~\m$ is determined.
\end{abstract}

\vspace{1.5cm}

\begin{center}
Submitted to \PLB 
\end{center}

\end{titlepage}

\newpage

\noindent
%-- H1AUTS Author list by names 
%-- Status: Wed Jun  1 14:13:45 CEST 2011  Number of authors = 203 
F.D.~Aaron$^{5,48}$,           %BUCH-PD        11/06           Aaron               
C.~Alexa$^{5}$,                %BUCH-PD        06/06           Alexa               
V.~Andreev$^{25}$,             %LPI -PD        8/88            Andreev             
S.~Backovic$^{30}$,            %PODG-PD        03/02           Backovic            
A.~Baghdasaryan$^{38}$,        %YERE-PD        09/03           Baghdasaryana       
S.~Baghdasaryan$^{38}$,        %YERE-ST        02/10           Baghdasaryans       
E.~Barrelet$^{29}$,            %PARI-PD        11/99           Barrelet            
W.~Bartel$^{11}$,              %DESY-PD        8/88            Bartel              
K.~Begzsuren$^{35}$,           %ULBA-PD        04/06           Begzsuren           
A.~Belousov$^{25}$
% ...
%-- H1 Institutes 

\noindent
\bigskip{\it \\
 $ ^{1}$ I. Physikalisches Institut der RWTH, Aachen, Germany \\
% ...
$ ^{5}$ National Institute for Physics and Nuclear Engineering (NIPNE) ,
          Bucharest, Romania$^{ m}$ \\
% ...
$ ^{11}$ DESY, Hamburg, Germany \\
% ...
$ ^{25}$ Lebedev Physical Institute, Moscow, Russia$^{ e}$ \\
% ...
$ ^{29}$ LPNHE, Universit\'e Pierre et Marie Curie Paris 6,
          Universit\'e Denis Diderot Paris 7, CNRS/IN2P3, Paris, France \\
 $ ^{30}$ Faculty of Science, University of Montenegro,
          Podgorica, Montenegro$^{ n}$ \\
% ...
$ ^{35}$ Institute of Physics and Technology of the Mongolian
          Academy of Sciences, Ulaanbaatar, Mongolia \\
% ...
$ ^{38}$ Yerevan Physics Institute, Yerevan, Armenia \\
% ...
$ ^{48}$ Also at Faculty of Physics, University of Bucharest,
          Bucharest, Romania \\
% ...
$ ^e$ Supported by the Deutsche Forschungsgemeinschaft \\
% ...
$ ^m$ Supported by the Romanian National Authority for Scientific Research
      under the contract PN 09370101 \\
 $ ^n$ Partially Supported by Ministry of Science of Montenegro,
      no. 05-1/3-3352 \\
}

\newpage

%\clearpage

\section{Introduction}

Deep inelastic neutral current (NC) scattering $e^\pm p \rightarrow e^\pm X$ at high 
negative four-momentum transfer squared \qsq allows the
structure of $e q$ interactions to be probed at short distances and to search for
new phenomena beyond the Standard Model (SM). 
Using the concept of four-fermion contact interactions (CI)
the interference of the photon and $Z$-boson fields with any new  
particle field associated to larger scales can be investigated.

Results from searches for contact interactions in $ep$ interactions at
HERA have been previously reported by the H1 \cite{h1ci2003,h1ci2000} and ZEUS 
\cite{zeusci2004,zeusci2000} collaborations. 
Therein, genuine contact interaction models, models with leptoquarks and 
supersymmetric scalar quarks (squarks),
low-scale quantum gravity models with large extra dimensions and
compositeness models
of quarks have been investigated by searching for deviations from
the SM expectation at high $Q^2$. Contact interaction studies have been also performed at LEP~\cite{LEP}.

Such models have also been investigated in direct searches at 
HERA, the Tevatron and the LHC.
Searches for leptoquarks involving lepton flavour violation \cite{h1lfv} and 
%scalar 
squarks in $R$-parity violating ($\rpv$) supersymmetric models \cite{h1rpv} 
have been published by the H1 collaboration using the full HERA data. 
Searches for leptoquark 
pair production in proton-proton 
collisions at a centre--of--mass energy of $\sqrt{s}=7$~TeV were reported by the 
ATLAS \cite{atlaslq} and CMS \cite{cmslq} collaborations, 
excluding first generation scalar leptoquarks up to $376$~TeV and $384$~TeV, 
respectively. 
These results surpass limits obtained at the Tevatron 
\cite{d0lq}. 
Stringent limits on low-scale quantum gravity models with large
extra dimensions using di-jet, di-electron and di-photon final states
have been reported by the D{\O} collaboration \cite{d0led2008,d0led2009} and 
recently,  by the CMS collaboration, excluding mass scales below $1.6$-$2.3$~TeV
depending on the model~\cite{cmsled}.

The analysis in this paper is based on the full H1 data sample collected in the years  1994-2007, which corresponds to an
integrated luminosity of $\L=446$~pb$^{-1}$ and represents a factor of $3(12)$ increase in statistics for  $e^+p$ ($e^-p$) collisions with respect to the previous publications \cite{h1ci2003}.
The same method is used as in previous analysis which is superseded by the results presented in this paper.
%%

%\cite{dummy}

\section{Contact Interaction Models}
New physics phenomena in fermion-fermion
scattering experiments may manifest themselves in deviations of the differential cross section 
\dsd {\qsq} from the SM expectation, and may be related to new heavy particles with masses $M_X$ 
much larger than the electroweak scale. 
In the low energy limit $\sqrt{s} \ll M_X$
such phenomena  can be described by an effective 
four-fermion CI model.
Different implementations of this effective model are summarised in the following.

\subsection{General contact interactions and compositeness}

In $ep$ scattering, the most general chiral invariant Lagrangian for neutral current vector-like  
four-fermion contact interactions can be written in the form~\cite{elpr,haberl}:
\begin{eqnarray}
  {\cal L}_V  &=& \sum_{q } 
  \sum_{a,\,b\, =\, L,\,R}
  \eta^q_{ab}\, (\bar{e}_a\gamma_\mu e_a)(\bar{q}_b\gamma^\mu q_b) \ ,
 \label{lcontact}
\end{eqnarray}
where $\eta^{q}_{ab}$ are the CI coupling coefficients, $a$ and $b$
indicate the left-handed and right-handed fermion helicities and the
first sum is over all quark flavours.
In the kinematic region of interest mainly the valence quarks ($u$ and
$d$) contribute.

In the case of general models of fermion compositeness or substructure
the CI coupling coefficients are defined as:
\begin{equation} 
\eta^q_{ab} = \epsilon^q_{ab}\, \frac{4\pi}{\Lambda^2} \ . 
\label{eq:composit}
\end{equation}
%
%Here the coupling strength is conventionally set to one \cite{elpr}.
%
New physics models are then characterised by a common compositeness
scale $\Lambda$ and the coefficients $\epsilon^q_{ab}$, which describe
the chiral structure of the coupling and may take the values $\pm 1$
or $0$, depending on the scenario, for example pure left-handed (L), right-handed (R),
or vector (V) and axial-vector (A) couplings.
Depending on the model and the sign of the coefficients, the new
physics processes interfere either constructively or destructively
with the SM processes.

\subsection{Leptoquarks}

Leptoquarks, colour triplet scalar or vector bosons carrying lepton
and baryon number, appear naturally in extensions of the SM which aim
to unify the lepton and quark sectors.
For leptoquark masses $M_{\rm LQ}$ much larger than the probing scale $M_{\rm LQ}
\gg \sqrt{s}$, the coupling $\lambda$ is related to the CI coupling coefficients via: 
\begin{equation}
  \eta^{q}_{ab} = \epsilon^{q}_{ab}\, \frac{\lambda^2}{M_{\rm LQ}^2} \ . 
\end{equation}
The classification of the leptoquarks follows the
Buchm\"uller-R\"uckl-Wyler (BRW) model~\cite{brw}, in which the
coefficients $\epsilon^{q}_{ab}$ depend on the leptoquark
type~\cite{kalinowski} and take values $0$, $\pm \frac{1}{2}$, $\pm
1$, $\pm 2$.

Two leptoquark types, $S_{0}^{L}$ and $\tilde{S}_{1/2}^{L}$, have quantum numbers
identical to %supersymmetric scalar 
the squarks $\tilde{d}$ and $\tilde{u}$.  
For these leptoquarks the couplings  $\lambda$  correspond to the
Yukawa couplings, $\lambda'_{ijk}$, 
which describe the ~$\rpv$ supersymmetric $L_iQ_j \bar{D}_k$ interaction \cite{rpv}.
Here $i$, $j$ and $k$ are the family indices and
$L_i$, $Q_j$ and $\bar{D}_k$ are the super-fields containing the left-handed leptons, 
the left-handed up-type quarks and the right-handed down-type quarks, 
respectively, together with their supersymmetric partners.

\subsection{Large extra dimensions}

In some string inspired models the small nature of the gravitational force is
explained by the existence of compactified extra dimensions~\cite{add}. 
In these models the gravitational scale %$M_S$ 
$M_{S}$ in $4+n$ dimensions is related
to the size $R$ of the compactified extra dimensions via the Planck scale 
%$M_P^2 \sim R^n\,M_S^{2+n}$.
$M_P^2 \sim R^n\,M_{S}^{2+n}$.
SM particles reside on a four-dimensional brane, while the spin $2$
graviton propagates into the extra spatial dimensions creating a tower
of Kaluza-Klein states.
Assuming that the ultraviolet cut-off scale of the tower is of similar
size to the gravitational scale, an effective contact-type
interaction~\cite{giudice} term can be defined with a coupling coefficient:
\begin{equation}
%  \eta_G = \frac{\lambda}{M^4_S} \  . 
  \eta_G = \frac{\lambda}{M^4_{S}} \  . 
\end{equation}
The coupling $\lambda$ depends on details of the theory and is
conventionally set to $\pm 1$.

\subsection{Quark Radius}
A clear manifestation of substructure would be the observation of
finite size effects like the measurement of electroweak charge distributions
of fermions.
Finite size 
effects are typically described by a standard form factor in the $eq$
scattering cross section:
\begin{eqnarray}
  f(Q^2) & = & 1 - \frac{{\langle R^2 \rangle}}{6}\,Q^2 \ , 
\end{eqnarray}
which relates the decrease of the scattering cross sections at high $Q^2$ to
 the mean squared radius $\langle R^2 \rangle$ of the electroweak charge distribution.
This form factor modifies the \qsq dependence of the $ep$ scattering cross
section similarly to the CI models described above.

\section{Data and Analysis Method} 

The analysed data sample is recorded in $e^+p$ and $e^-p$ collisions corresponding to an
integrated luminosity of $281$~pb$^{-1}$ and $165$~pb$^{-1}$, respectively.
The measurement of the differential neutral current cross-section, \dsd {\qsq},
which is used to probe possible CI signatures %is a repetition of 
follows the previous measurements based on data
recorded in the years 1994-2000 \cite{h1xsec,h1xe+p,h1xe-p} and includes
new data recorded from 2003-2007. 
A list of the analysed data sets is given in table~\ref{datasets}.

\begin{table}[htb]
\begin{center} %\vspace*{-2mm}
  \begin{tabular}{cccl@{\,}c}
    \toprule
          Reaction & $\L_{int}~[\pb^{-1}]$ & $\sqrt{s}~[\GeV]$ 
        & \multicolumn{2}{c}{Polarisation $(P_e~[\%])$}  \\[.2ex]
        \midrule
          $e^+p \to e^+X$ & $36$ & $301$
        & \multicolumn{2}{c}{Unpolarised}   \\[.2ex]
          $e^-p \to e^-X$ & $16$ & $319$
        & \multicolumn{2}{c}{Unpolarised}   \\[.2ex]
          $e^+p \to e^+X$ & $65$ & $319$
        & \multicolumn{2}{c}{Unpolarised}  \\[.2ex]
          $e^-p \to e^-X$ & $46$ & $319$
        & Right & $(P_e=+37)$  \\[.2ex]
          $e^-p \to e^-X$ & $103$ & $319$
        & Left  & $(P_e=-26)$  \\[.2ex]
          $e^+p \to e^+X$ & $98$ & $319$
        & Right & $(P_e=+33)$  \\[.2ex]
          $e^+p \to e^+X$ & $82$ & $319$
        & Left  & $(P_e=-38)$  \\
        \bottomrule
\end{tabular}
\end{center}
\caption{%H1 
Data samples recorded in the years 1994-2007 with corresponding
	integrated luminosities, centre-of-mass energies and average longitudinal
	polarisations.}
\label{datasets}
\end{table}

The data collected from the year 2003 onwards were taken 
with a longitudinally
polarised lepton beam, with typical polarisation values of
$\pm 35\%$. 
The average luminosity weighted polarisations of the
$e^+p$ and $e^-p$ data sets are small.
Due to the different chiral structure of
hypothetical new particles with respect to the irreducible
SM background, the sensitivity 
to possible new physics phenomena is increased by up to $15$ percent by 
analysing the data sets with left and right 
longitudinal lepton polarisation separately.

Contact interactions are investigated by searching for
deviations in the NC differential cross section \dsd{Q^2} from the SM
expectation at high negative four-momentum transfer squared $Q^2>200$~GeV$^2$. 
The  SM cross section of neutral current scattering
factorises into the electroweak matrix element of the hard 
$eq$ interaction process and 
the parton distribution function (PDF) of the proton. 
The \qsq dependence of the PDF is calculated using perturbative QCD
\cite{DGLAP}. 

For the CI analysis the parton
densities at high values of $Q^2$, corresponding to high values of $x$, are of
special importance.
In this analysis the CTEQ6m~\cite{cteq6} PDF is used to calculate both the 
SM and signal expectations.
The CTEQ6m set was obtained by fitting several experimental
data sets.
At high $x$ %, the region relevant for this analysis, 
this PDF is mostly constrained by fixed target experiments and also 
by $W$-boson production and jet data from the Tevatron experiments,
which are not sensitive to possible $eq$ contact interaction processes. %signals. 
CTEQ6m also includes early $e^\pm p$  scattering data at high \qsq from the
H1~({$\L=52~\textrm{pb}^{-1}$) and ZEUS~({$ \L=30~\textrm{pb}^{-1}$) experiments. 
However, since the $e^+ p$ ($e^- p$) data sets analysed here are $6(10)$
times larger, the residual correlations between the HERA data and the 
CTEQ6m PDF are small and are neglected in the following.
Furthermore, the CTEQ6m parton densities can be regarded as unbiased with 
respect to possible contact interaction effects.
 CTEQ6m is chosen as it describes many
experimental data and in particular, the HERA data in the region
\qsq$<200$~GeV$^2$, which are not used in this analysis.
The results of this analysis are
verified using an alternative PDF not based 
on HERA high \qsq data, as described in section~\ref{sec:results}.

The single differential cross sections \dsd{\qsq} are measured
for $e^+p$ and $e^-p$ scattering up to \qsq$=30000$~GeV$^2$
and compared to the SM expectation.
The ratio of the data to the SM expectation is shown in figure~\ref{fig:dsdq2norm}.
Good agreement between data and the SM is observed,
in particular in the high \qsq region, which is the focus of this analysis.

In the next step, a quantitative test of the SM and the CI models is performed by
investigating the measured cross sections ${\rm d}\sigma / {\rm d}Q^2$ 
following the analysis method described in \cite{h1ci2003} by
applying a minimisation of the $\chi^2$ function \cite{h1lowx}:
\begin{equation}
  \chi^2(\eta,\varepsilon) = \sum_i \frac {\left( {\sigma}_i^{\rm exp}-
      {\sigma}_i^{\rm th}(\eta)\, (1-\sum_k \Delta_{ik}(\varepsilon_k))  \right)^2}
      {\delta^{2}_{i,\rm{stat}}\,{\sigma}_i^{\rm exp}{\sigma}_i^{\rm th}(\eta)\,
(1-\sum_k \Delta_{ik}(\varepsilon_k)) +  {(\delta_{i,\rm{uncor}}\,{\sigma}_i^{\rm exp})}^2}
+ \sum_k \varepsilon_k^2 \ . 
  \label{chi2}
\end{equation}
Here ${\sigma}_i^{\rm exp}$ and ${\sigma}_i^{\rm th}(\eta)$ are the
experimental and theoretical cross sections, respectively, for the measurement point $i$, and
${\delta}_{i,\rm{stat}}$ and ${\delta}_{i,\rm{uncor}}$ correspond to the relative statistical
and uncorrelated systematic errors, respectively.
The theoretical cross section includes both the SM and the contact interaction
term, 
and it depends on the coupling
coefficient $\eta$, which is varied in the fit.
The functions $\Delta_{ik}(\varepsilon_k)$ describe the correlated systematic
errors for point $i$ associated to a source $k$ and
depend on the fit parameters $\varepsilon_k$.
In particular, the normalisations of the individual data sets described in table~\ref{datasets}
are free parameters, only constrained by the individual luminosity 
measurements.
Since the precise cross section measurements at low \qsq determine the normalisations,
new physics signals are mainly tested by exploiting 
the shape of the \qsq distribution.

Statistical and systematic uncertainties are taken
into account in the fit procedure.
The following sources of experimental uncertainties are accounted for 
\cite{h1xsec,h1xe-p,h1xe+p,theses}:
the electromagnetic energy scale uncertainty of $1-3\%$, 
the polar angle uncertainty of the scattered lepton of $2-3$~mrad, 
the uncertainty on the electron identification of $0.5-2\%$, 
the hadronic energy scale uncertainty of $2-7\%$
the uncertainty from the luminosity measurement of $1.6-3.8\%$ and
the uncertainty on the electron beam polarisation of $1-2.3\%$.
The effect of the above systematic uncertainties on the SM expectation is
determined by varying the experimental quantities by $\pm 1$ standard deviation 
in the MC samples and propagating these variations through the whole analysis.
The uncorrelated systematic uncertainties of the measurements 
vary as function of $Q^2$ between $1-11\%$ $(1.6-13\%)$ for
$e^+p$ ($e^-p$) scattering. 
The dominant sources of the correlated systematic errors are
the PDF uncertainty (about $8\%$), %and 
the uncertainties from the luminosity
measurement and from the experimental uncertainties on the energy scale and the polar angle of the
scattered lepton. 
%In addition, also the experimental uncertainties of up to $3\%$ each from the energy scale and from
%the polar angle measurement contribute to the correlated systematic error. 
All other experimental systematic uncertainties are found to have a negligible 
impact on the analysis.

%E 2.6  theta 3.4 
%\bf (?? -- add the correlated part) \rm

%All statistical and systematic uncertainties are taken
%into account in the fitting procedure.
%The dominant experimental error comes from the electromagnetic 
%energy scale uncertainty with some
%smaller contribution from the measurement of the electron polar angle.
%Other experimental uncertainties coming from the lepton identification and 
%the trigger employed in the analysis are negligible.

% uncorrelated            10.85%        13.01%
%Electron energy(corr)        1.33%        0.51%
%Electron theta(corr)        0.26%        0.46%
%PDF(corr)            7.6%        8.4%
%lumi                3.1%        3.8%

\section{Results} \label{sec:results}
The data used in this analysis
are found to be consistent with the expectation from the SM alone
($\eta=0$ in equation~\ref{chi2}) based on the CTEQ6m PDF, yielding a $\chi^2/dof=16.4/17$ $(7.0/17)$ 
for the $e^+p$ $(e^-p)$ data.
The normalisation constants of the
individual data sets agree well with the SM expectation within the PDF
uncertainties.

For each CI model the
effective scale parameters and couplings
describing the new physics scale
are determined by a fit to the 
differential NC cross section.
All scale parameters are
found to be consistent with the SM and limits are calculated at $95\%$
confidence levels (CL) using the frequentist method as
described in the previous publication \cite{h1ci2003}.

Lower limits on the compositeness scale $\Lambda$ in the context
of the general contact interaction model are presented in table~\ref{etafits}
and figure~\ref{fig:lqlimits}. 
The results are presented for eight scenarios, which differ in their
chiral structure as determined by the CI coupling coefficients
$\eta^{q}_{ab}$.
%The limits are given for both signs of coupling constants, where 
%$\Lambda^+$ ($\Lambda^-$) denotes the positive (negative) sign.
%
Depending on the model and the sign of the coefficients,
limits on $\Lambda$ are obtained in the range $3.6~\TeV$  to $7.2~\TeV$. 
In figure~\ref{fig:dsdq2normvv}, differential cross section
measurements for $e^+p$ and $e^-p$ scattering normalised to the SM
expectation are compared to the predictions corresponding to the
$95$\% CL exclusion limits of the VV model, $\Lambda^+_{VV}>5.6~\TeV$
and $\Lambda^-_{VV}>7.2~\TeV$.

For leptoquark-type contact interactions, the notation, 
quantum numbers and lower limits on $M_{\rm LQ}/\lambda$ 
are presented in table~\ref{lqfits}.
The limits are in the range $M_{\rm LQ}/\lambda > 0.41-1.86~\TeV$. 
Leptoquarks coupling to $u$ quarks are probed with higher 
sensitivity, corresponding to more stringent limits than those
coupling to $d$ quarks due to the different quark densities
in the proton.
In figure~\ref{fig:dsdq2normlq}, 
the normalised differential $e^\pm p$ cross section measurements 
are compared to the predicted cross sections corresponding to the $95\%$ CL exclusion limits 
of the $S_1^L$ and $V_1^L$ leptoquarks. 
At high \qsq the existence of a $S_1^L$ ($V_1^L$) leptoquark would lead to an increase
(decrease) of the $e^\pm p$ cross sections, which is not observed. 
For a Yukawa coupling of electromagnetic strength, $\lambda=0.3$, 
scalar and vector leptoquark masses up to $0.33~\TeV$ and $0.56~\TeV$ are
excluded, respectively,
comparable or exceeding limits obtained by Tevatron and LHC. 
The leptoquarks $S_0^L$ and $\tilde{S}_{1/2}^L$ may also be  
interpreted as squarks in the framework of $\rpv$ supersymmetry and the
corresponding limits in terms of the ratio $M_{\tilde{q}}/\lambda'$ 
are given in table~\ref{sqfits}.
For a Yukawa coupling of electromagnetic strength, the corresponding
lower limit on the $\tilde{u}$ mass of $0.33$~TeV is similar to that
obtained recently by the H1 collaboration in a direct search \cite{h1rpv}. 
%%
%Schael 2007, s~_R m>490 GeV

Lower limits in a model with large extra dimensions on the
gravitational scale $M_{S}$ in $4+n$~dimensions assuming a positive ($\lambda = +1$) or
negative ($\lambda = -1$) coupling are given in
table~\ref{ledfits}.
Mass scales $M_{S}<0.9~\gev$ are excluded at $95\%$ CL.
The corresponding cross section predictions normalised to the SM expectation are compared to the $e^{\pm}p$ data in figure~\ref{fig:dsdq2normled}.

Finally, an upper limit at $95\%$~CL on the quark radius $R_q < 0.65 \cdot
10^{-18}$~m is derived assuming point-like leptons. 
The corresponding cross section predictions normalised to the SM expectation
are compared to the $e^{\pm}p$ data in figure~\ref{fig:dsdq2normrad}.
%%

% Voica
%The above results were also verified using the dedicated H1 PDF, where
%a good agreement was found between results based on the dedicated H1 PDF set
%and the CTEQ6m PDF.

The above results are also verified  using a dedicated H1 PDF set
based on data collected in the years 1994-2007.
This PDF set was  obtained from a next-to-leading order QCD fit to the H1 data \cite{h1lowx}
with $Q^2<200$~GeV$^2$, excluding the high \qsq data used in
this analysis.
Both the SM expectation and limits derived using the dedicated H1 PDF
agree well with those obtained using the CTEQ6m PDF within the uncertainties.

\section{Summary}

Neutral current deep inelastic $e^-p$ and $e^+p$ scattering
cross section measurements
are analysed to search for new phenomena mediated via contact interactions.
The data are well described by the Standard Model expectations.
Limits on the %strength 
parameters of various contact interaction models are presented at $95\%$ CL.

Lower limits on the compositeness scale $\Lambda$ are derived within a general
contact interaction analysis.
The limits range between $3.6~\TeV$ and $7.2~\TeV$ depending on the chiral
structure, corresponding to an increase by a factor of about two  
compared to previous HERA searches. 
The study of leptoquark exchange yields lower limits on 
the ratio $M_{\rm LQ}/\lambda$ between $0.41~\TeV$ and $1.86~\TeV$,
considerably improving constraints from the previous analysis. 
Squarks in the framework of $R$-parity violating supersymmetry 
with masses satisfying
$M_{\tilde u}/\lambda'_{1j1} < 1.10~\TeV$ and
$M_{\tilde d}/\lambda'_{11k} < 0.66~\TeV$ are excluded.
Possible effects of low-scale quantum gravity with gravitons 
propagating into extra spatial dimensions are also investigated, where
lower limits %on the effective Planck 
on the gravitational scale in $4+n$~dimensions $M_{S}>0.9~\TeV$
are found.
Finally, a form factor approach yields an upper limit on the size
of light $u$ and $d$ quarks of $R_q< 0.65\cdot 10^{-18}~\m$,
assuming point-like leptons.

Using the full HERA data set, limits derived in this analysis  
are more stringent than previous results by H1 and ZEUS.
The results can also be compared to those obtained 
by the LEP, Tevatron and, most recently, LHC collaborations.
For most models with Yukawa couplings of
electromagnetic strength, or stronger,
the analysis presented here provides the most stringent limits on
first generation leptoquarks. 

\section*{Acknowledgements}
We are grateful to the HERA machine group whose outstanding 
efforts have made this experiment possible. 
We thank the engineers and technicians for their work in constructing and
maintaining the H1 detector, our funding agencies for financial support, 
the DESY technical staff for continual assistance and the DESY directorate 
for support and for the hospitality which
they extend to the non-DESY members of the collaboration.

\clearpage

%
%   References for Contact Interaction paper
%

\clearpage

% compositeness scales
\begin{table}[htb]
\begin{center}
\begin{tabular}{l c c c}
%   \hdick \\[-1.5ex]
    \hline \\[-1.5ex]
          %& \multicolumn{2}{c}{$e^-p$ $(319\,\GeV)$}
          %& \multicolumn{2}{c}{$e^+p$ $(319\,\GeV)$}
          \multicolumn{4}{c}{\bf\large H1 Search for General Compositeness} \\ [1ex]
         % \multicolumn{4}{c}{$\eta^{eq}_{ij} = \epsilon^{q}_{ij}4\pi/\Lambda^{2}$} \\
   %\hdick \\[-3.ex]
          & $\eta^{q}_{ab} = \epsilon^{q}_{ab}~4\pi/\Lambda^{2}$  
          & \multicolumn{2}{c}{} \\ [0.6ex]
          Model    %& \ $\Lambda^+~[\TeV]$  &  $\Lambda^-~[\TeV]$ \
          %& \ $\Lambda^+~[\TeV]$  &  $\Lambda^-~[\TeV]$ \ \
          & [\tf{$\epsilon_{LL}$},\tf{$\epsilon_{LR}$}, \tf{$\epsilon_{RL}$}, \tf{$ \epsilon_{RR}$}]
          & $\Lambda^+~[\TeV]$  &  $\Lambda^-~[\TeV]$ \\[1ex]
%   \hdick \\[-1.5ex]
    \hline \\[-1.5ex]
 $LL$  & [\tf{$\pm 1$},\tf{$0$},\tf{$0$},\tf{$0$}] & $4.2$ & $4.0$  \\[.2em]
 $LR$  & [\tf{$0$},\tf{$\pm 1$},\tf{$0$},\tf{$0$}] & $4.8$ & $3.7$  \\[.2em]
 $RL$  & [\tf{$0$},\tf{$0$},\tf{$\pm1$},\tf{$0$}] & $4.8$ & $3.8$  \\[.2em]
 $RR$  & [\tf{$0$},\tf{$0$},\tf{$0$},\tf{$\pm1$}] & $4.4$ & $3.9$  \\[0.6ex]
 $VV$  & [\tf{$\pm 1$},\tf{$\pm 1$},\tf{$\pm 1$},\tf{$\pm 1$}] & $5.6$ & $7.2$  \\[.2em]
 $AA$  & [\tf{$\pm 1$},\tf{$\mp 1$},\tf{$\mp 1$},\tf{$\pm 1$}] & $4.4$ & $5.1$  \\[.2em]
 $VA$  & [\tf{$\pm 1$},\tf{$\mp 1$},\tf{$\pm 1$},\tf{$\mp 1$}] & $3.8$ & $3.6$  \\[0.6ex]
% $X1$  & [\tf{$\pm1$},\tf{$\pm 1$},\tf{$0$},\tf{$0$}] & $3.3$ & $4.5$  \\[.2em]
% $X2$  & [\tf{\pm1$},\tf{0$},\tf{\pm1$},\tf{0$}] & 5.1 & 5.0  \\[.2em]
 $LL+RR$  & [\tf{$\pm 1$},\tf{$0$},\tf{$0$},\tf{$\pm 1$}] & $5.3$ & $5.1$  \\[.2em]
 $LR+RL$  & [\tf{$0$},\tf{$\pm 1$},\tf{$\pm1$},\tf{$0$}] & $5.4$ & $4.8$  \\[.2em]
% $X5$  & [\tf{0},\tf{+1},\tf{0},\tf{+1}] & 5.1 & 5.0  \\[.2em]
% $X6$  & [\tf{0},\tf{0},\tf{+1},\tf{$-1$}] & 4.1 & 3.2  \\[.2em]
% $U1$  & [\tf{+1},\tf{$-1$},\tf{0},\tf{0}] & 4.0 & 4.7  \\[.2em]
% $U2$  & [\tf{+1},\tf{0},\tf{+1},\tf{0}] & 5.2 & 5.5  \\[.2em]
% $U3$  & [\tf{0},\tf{+1},\tf{0},\tf{+1}] & 5.2 & 7.1  \\[.2em]
% $U4$  & [\tf{0},\tf{+1},\tf{+1},\tf{0}] & 5.3 & 6.4  \\[.2em]
% $U5$  & [\tf{0},\tf{+1},\tf{0},\tf{+1}] & 5.2 & 5.8  \\[.2em]
% $U6$  & [\tf{0},\tf{0},\tf{+1},\tf{$-1$}] & 4.1 & 3.5  \\[.2em]
\hline
\end{tabular}
\end{center}

\caption{Lower limits at $95\%$~CL on the compositeness scale $\Lambda$. The $\Lambda^{+}$
	limits correspond to the upper signs and the $\Lambda^{-}$ limits correspond to
	the lower signs of the chiral coefficients
	\mbox{[$\epsilon^q_{LL}$, $\epsilon^q_{LR}$, $\epsilon^q_{RL}$, $\epsilon^q_{RR}$]}.}
\label{etafits}
\end{table}

\clearpage

% leptoquarks
\begin{table}[htb]
%\label{tab:LQ}
\begin{center}
\begin{tabular}{l c c c c c}
%  \hdick \\[-1.5ex]
   \hline \\[-1.5ex]
     \multicolumn{5}{c}{\bf\large H1 Search for Heavy Leptoquarks} \\ [1ex]
    % \hdick \\[-1.5ex]
    & \multicolumn{2}{c}{$\eta^q_{ab}=\epsilon^q_{ab}~\lambda^2 /M_{\rm LQ}^2$ }
    &  %& $e^-p$ $(319~\GeV)$ & $e^+p$ $(319~\GeV)$
    & \\[.5ex]
     LQ & $\epsilon^u_{ab}$ & $\epsilon^d_{ab}$
    & $F$ %& $M_{\rm LQ}/\lambda ~ [ \GeV ]$
          %& $M_{\rm LQ}/\lambda ~ [ \GeV ]$
          & $M_{\rm LQ}/\lambda ~ [ \TeV ]$ \\[1ex]
%  \hdick \\[-1.5ex]
   \hline \\[-1.5ex]
   $S_0^L$ &
    \ $\epsilon^u_{LL} = +\frac{1}{2}$ \ & & $2$ &  $1.10$ \\[.2em]
   $S_0^R$ &
    \ $\epsilon^u_{RR} = +\frac{1}{2}$ \ & & $2$ &  $1.10$ \\[.2em]
   $\tilde{S}_0^R$ & &
    \ $\epsilon^d_{RR} = +\frac{1}{2}$ \   & $2$ &  $0.41$ \\[.2em]
   $S_{1/2}^L$ &
    \ $\epsilon^u_{LR} = -\frac{1}{2}$ \ & & $0$ & $0.87$ \\[.2em]
   $S_{1/2}^R$ &
    \ $\epsilon^u_{RL} = -\frac{1}{2}$ \   &
    \ $\epsilon^d_{RL} = -\frac{1}{2}$ \   & $0$ &  $0.59$ \\[.2em]
   $\tilde{S}_{1/2}^L$ & &
    \ $\epsilon^d_{LR} = -\frac{1}{2}$ \   & $0$ &  $0.66$ \\[.2em]
   $S_1^L$ &
    \ $\epsilon^u_{LL} = +\frac{1}{2}$ \   &
    \ $\epsilon^d_{LL} = +1$ \             & $2$ &  $0.71$ \\[1ex]
  \hline \\[-1.5ex]
   $V_0^L$ & &
    \ $\epsilon^d_{LL} = -1$ \   & $0$ &  $1.06$ \\[.2em]
   $V_0^R$ & &
    \ $\epsilon^d_{RR} = -1$ \   & $0$ &  $0.91$ \\[.2em]
   $\tilde{V}_0^R$ &
    \ $\epsilon^u_{RR} = -1$ \ & & $0$ & $1.35$ \\[.2em]
   $V_{1/2}^L$ & &
    \ $\epsilon^d_{LR} = +1$ \   & $2$ &  $0.51$ \\[.2em]
   $V_{1/2}^R$ &
    \ $\epsilon^u_{RL} = +1$ \ &
    \ $\epsilon^d_{RL} = +1$ \   & $2$ & $1.44$ \\[.2em]
   $\tilde{V}_{1/2}^L$ &
    \ $\epsilon^u_{LR} = +1$ \ & & $2$ & $1.58$ \\[.2em]
   $V_1^L$ &
    \ $\epsilon^u_{LL} = -2$ \ &
    \ $\epsilon^d_{LL} = -1$ \   & $0$ & $1.86$ \\[1ex]
  \hline
\end{tabular}
\end{center}
\caption{Lower limits at $95\%$~CL on $M_{\rm LQ}/\lambda$ for scalar ($S$) and vector ($V$)
  	leptoquarks, where $L$ and $R$ denote the lepton chirality and the subscript
	($0,\ 1/2,\ 1$) is the weak isospin. For each leptoquark type, the relevant coefficients
	$\epsilon^{q}_{ab}$ and fermion number $F = L + 3B$ are indicated. Leptoquarks with
	identical quantum numbers except for weak hypercharge are distinguished using a tilde,
	for example $V_0^R$ and $\tilde{V}_0^R$. Quantum numbers and helicities refer to
	$e^-q$ and $e^-\bar{q}$ states.}
\label{lqfits}
\end{table}

\vspace{1cm}

% squarks
\begin{table}[htb]
\begin{center}
\begin{tabular}{lccc}
%  \hdick \\[-1.5ex]
  \hline \\[-1.5ex]
    \multicolumn{4}{c}{\bf\large H1 Search for {\boldmath$\brpv$} Squarks } \\ [1ex]
    %\hdick \\[-1.5ex]
%    &         %& $e^-p$ $(319~\GeV)$ & $e^+p$ $(319~\GeV)$
%    &  \\[.5ex]
    \ Channel & Coupling & $\epsilon^q_{ab}$
    %& $M_{\tilde q}/\lambda' ~ [ \GeV ]$
    %& $M_{\tilde q}/\lambda' ~ [ \GeV ]$
    & $M_{\tilde q}/\lambda' ~ [ \TeV ]$\\[1ex]
%  \hdick \\[-1.5ex]
   \hline \\[-1.5ex]
   $e^+ d \to \tilde u^{\,(k)}$ & $\lambda'_{11k}$ &
    \ $\epsilon^u_{LL} = +\frac{1}{2}$ \ & $1.10$ \\[.2em]
   $e^- u \to \tilde{d}^{\,(j)}$ & $\lambda'_{1j1}$ &
    \ $\epsilon^d_{LR} = -\frac{1}{2}$ \   & $0.66$ \\[1ex]
  \hline
\end{tabular}
\end{center}
\caption{Lower limits at $95\%$~CL on $M_{\tilde q}/\lambda'$ for the $R_p$ violating
	couplings $\lambda'_{ijk}$~\cite{rpv}, where $i,j,k$ are family indices.
	The coefficients $\epsilon^q_{ab}$ are also shown. The $\lambda'_{11k}$ ($\lambda'_{1j1}$)
	coupling corresponds to the $S_{0}^{L}$ ($\tilde{S}_{1/2}^{L}$) leptoquark coupling
	shown in table~\ref{lqfits}.}
\label{sqfits}
\end{table}

% large extra dimensions
\begin{table}[ht]
\begin{center}
\begin{tabular}{p{0.165\textwidth}cp{0.2\textwidth}}
%   \hdick \\[-1.5ex]
  \hline \\[-1.5ex]
          \multicolumn{3}{c}{\bf\large H1 Search for Large Extra Dimensions} \\ [1ex]
          %\hdick \\[-1.5ex]
          %& \ $e^- p$ $(319~\GeV)$ \ & \ $e^+ p$ $(319~\GeV)$ \
          & $\eta_{G} = \lambda/M_{S}^{4}$\\
 \multicolumn{1}{c}{coupling $\lambda$}  &  & \multicolumn{1}{c}{$M_{S}~[\TeV]$}  %& \ $M_S~[\TeV]$ \  & \ $M_S~[\TeV]$ \
  \\[1ex]
 % \hdick \\[-1.5ex]
  \hline \\[-1.5ex]
 \multicolumn{1}{c}{$+1$} &  & \multicolumn{1}{c}{$0.90$} \\[.2em]
 \multicolumn{1}{c}{$-1$} &  & \multicolumn{1}{c}{$0.92$} \\[.2em]
 \hline
\end{tabular}
\end{center}
\caption{Lower limits at $95\%$~CL on a model with large extra dimensions on the gravitational scale
	$M_{S}$ in $4+n$ dimensions, assuming positive ($\lambda = +1$) or negative ($\lambda = -1$) couplings.}
\label{ledfits}
\end{table}

\vspace{1cm}

\vfill
\clearpage

\begin{figure}
\includegraphics[width=\textwidth]{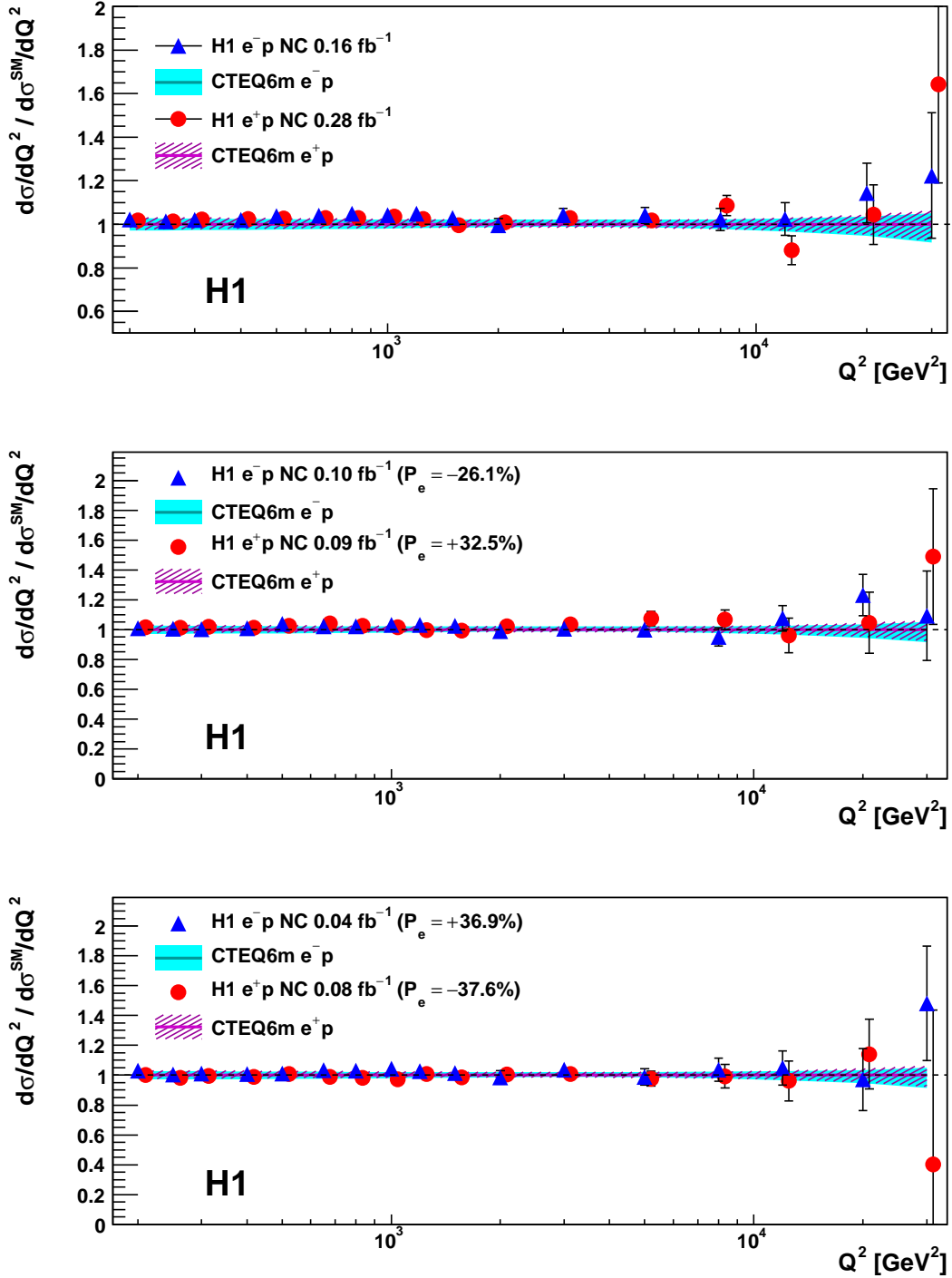}
\caption{The ratio of the measured cross section to the Standard Model prediction
	determined using the CTEQ6m PDF set for $e^+p \rightarrow
	e^+X$ and $e^-p \rightarrow e^-X$ scattering. The top figure%plot
	corresponds to the full H1 data with an average longitudinal
	polarisation of $P \approx 0$.  The middle and bottom figures%plots
	represent polarised H1 data taken from the year 2003 onwards
	for different lepton charge and polarisation data sets. The
	error bars represent the statistical and uncorrelated
	systematic errors added in quadrature. The bands indicate the
	PDF uncertainties of the Standard Model cross section
	predictions.}
\label{fig:dsdq2norm}
\end{figure}

\begin{figure}
\includegraphics[width=\textwidth]{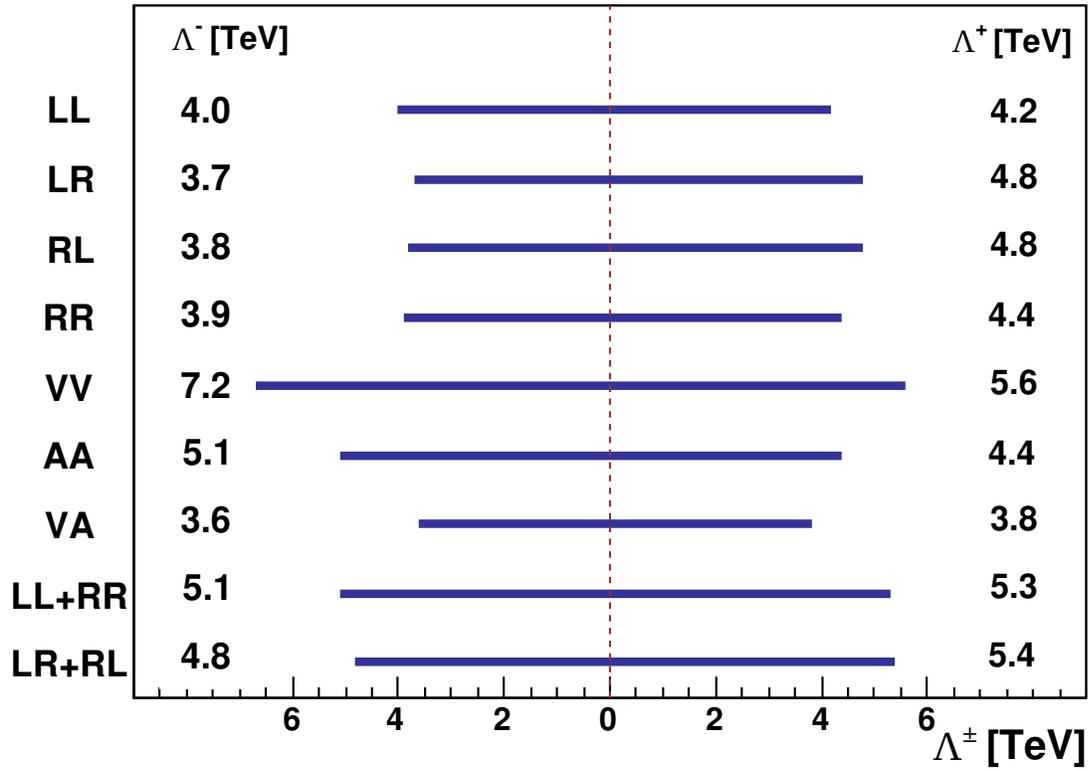}
\caption{Lower limits at $95\%$~CL on the compositeness scale $\Lambda$ for
	various chiral models, obtained from the full H1 data. Limits
	are given for both signs $\Lambda^+$ and $\Lambda^-$ of the chiral
	coefficients.}
\label{fig:lqlimits}
\end{figure}

\begin{figure}
\includegraphics[width=\textwidth]{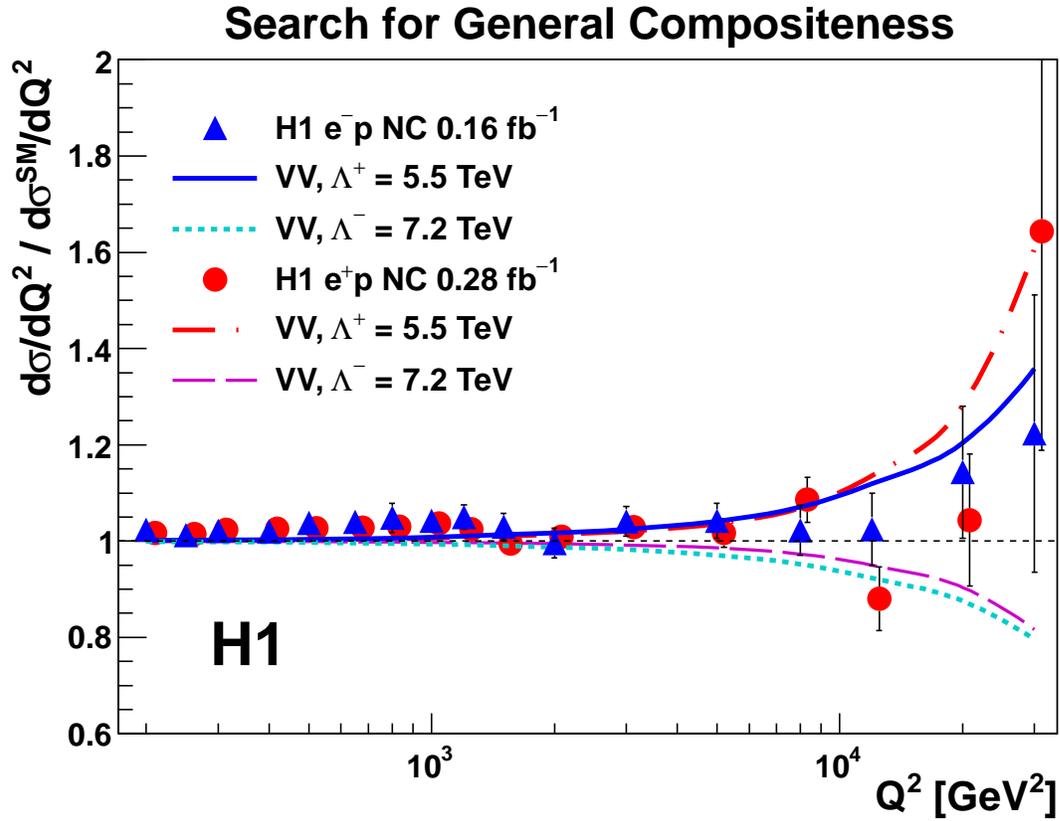}
\caption{The measured neutral current cross section ${\rm d}\sigma/{\rm d}Q^2$
	normalised to the Standard Model expectation. H1 $e^{\pm}p$ scattering
	data are compared with curves corresponding to $95$\%~CL exclusion
	limits obtained from the full H1 data for the $VV$ compositeness
	scale model, for both signs $\Lambda^+$ and $\Lambda^-$ of the chiral
	coefficients. The error bars represent the statistical and uncorrelated
	systematic errors added in quadrature.}
\label{fig:dsdq2normvv}
\end{figure}

\begin{figure}
\includegraphics[width=\textwidth]{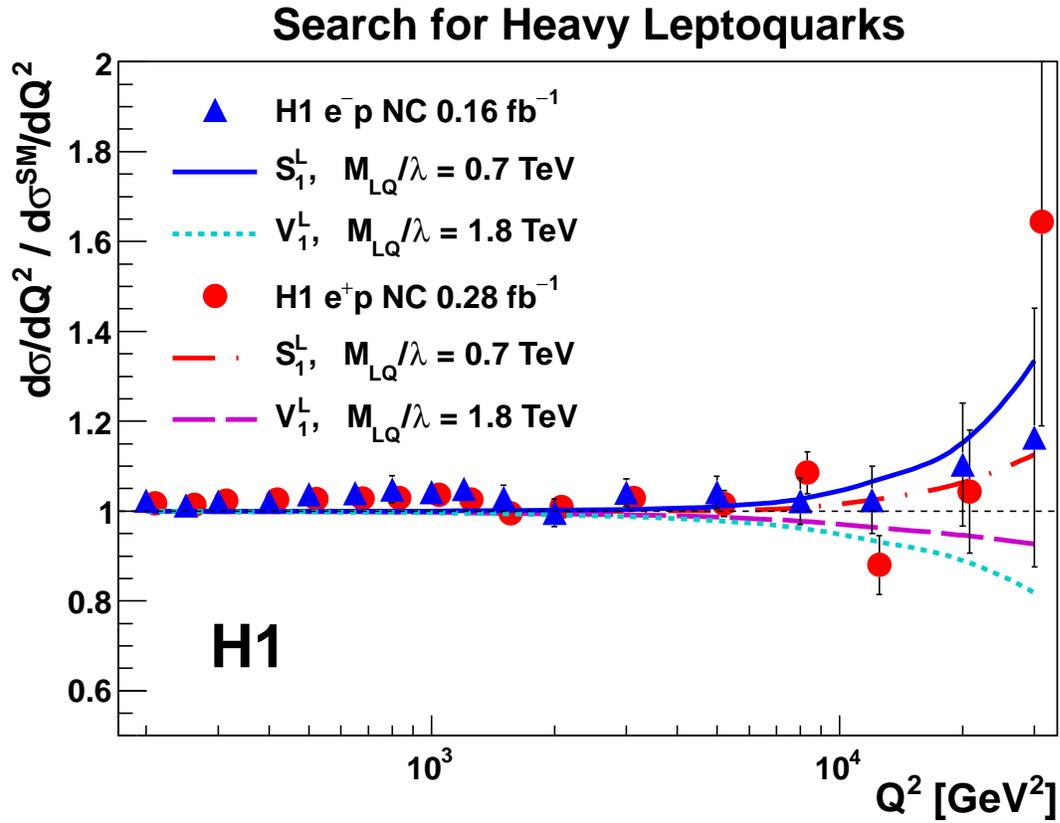}
\caption{The measured neutral current  cross section ${\rm d}\sigma/{\rm d}Q^2$ 
	normalised to the Standard Model expectation. H1 $e^{\pm}p$ scattering
	data are compared with curves corresponding to $95\%$~CL exclusion
	limits obtained from the full H1 data on the ratio
	$M_{\rm LQ}/\lambda$ for the $S^L_{1}$ and $V^L_{1}$ leptoquarks.
	The error bars represent the statistical and uncorrelated systematic
	errors added in quadrature.}
\label{fig:dsdq2normlq}
\end{figure}

\begin{figure}
\includegraphics[width=\textwidth]{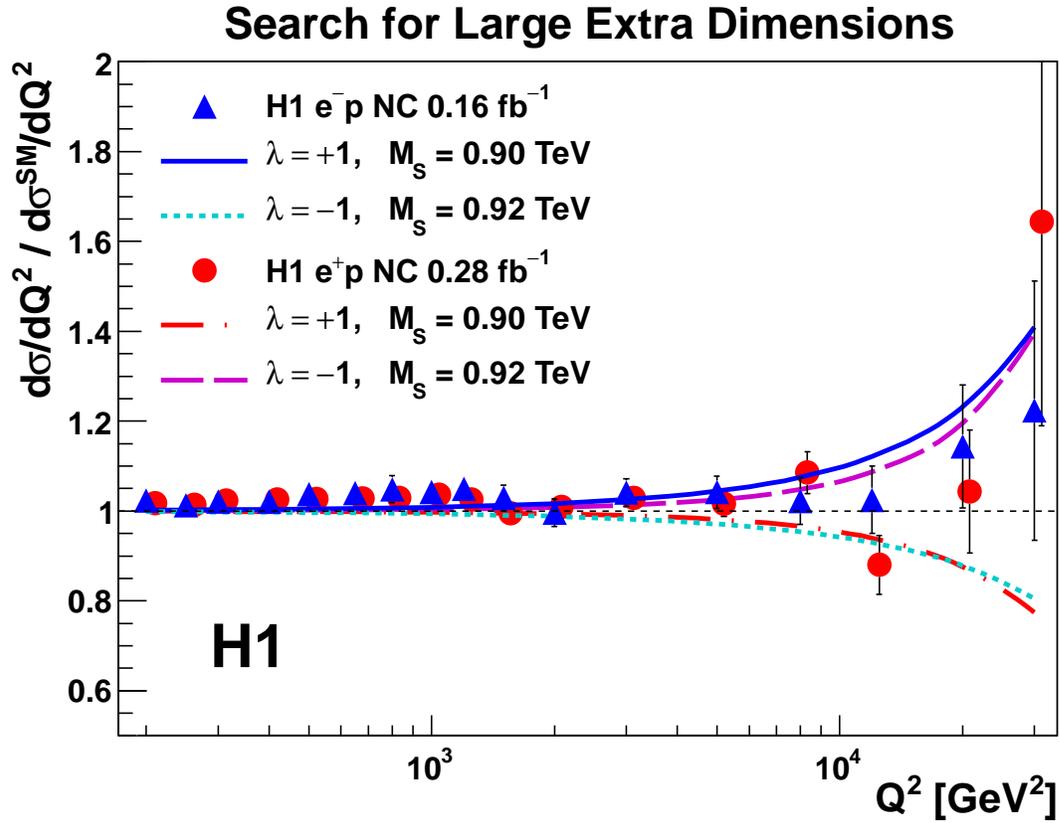}
\caption{The measured neutral current cross section ${\rm d}\sigma/{\rm d}Q^2$
	normalised to the Standard Model expectation. H1 $e^{\pm}p$ scattering
	data are compared with curves corresponding to $95\%$~CL exclusion
	limits obtained from the full H1 data on the gravitational scale,
	$M_{S}$ for both positive ($\lambda = +1$) and negative ($\lambda = -1$)
	couplings. The error bars represent the statistical and uncorrelated
	systematic errors added in quadrature.}
\label{fig:dsdq2normled}
\end{figure}

\begin{figure}
\includegraphics[width=\textwidth]{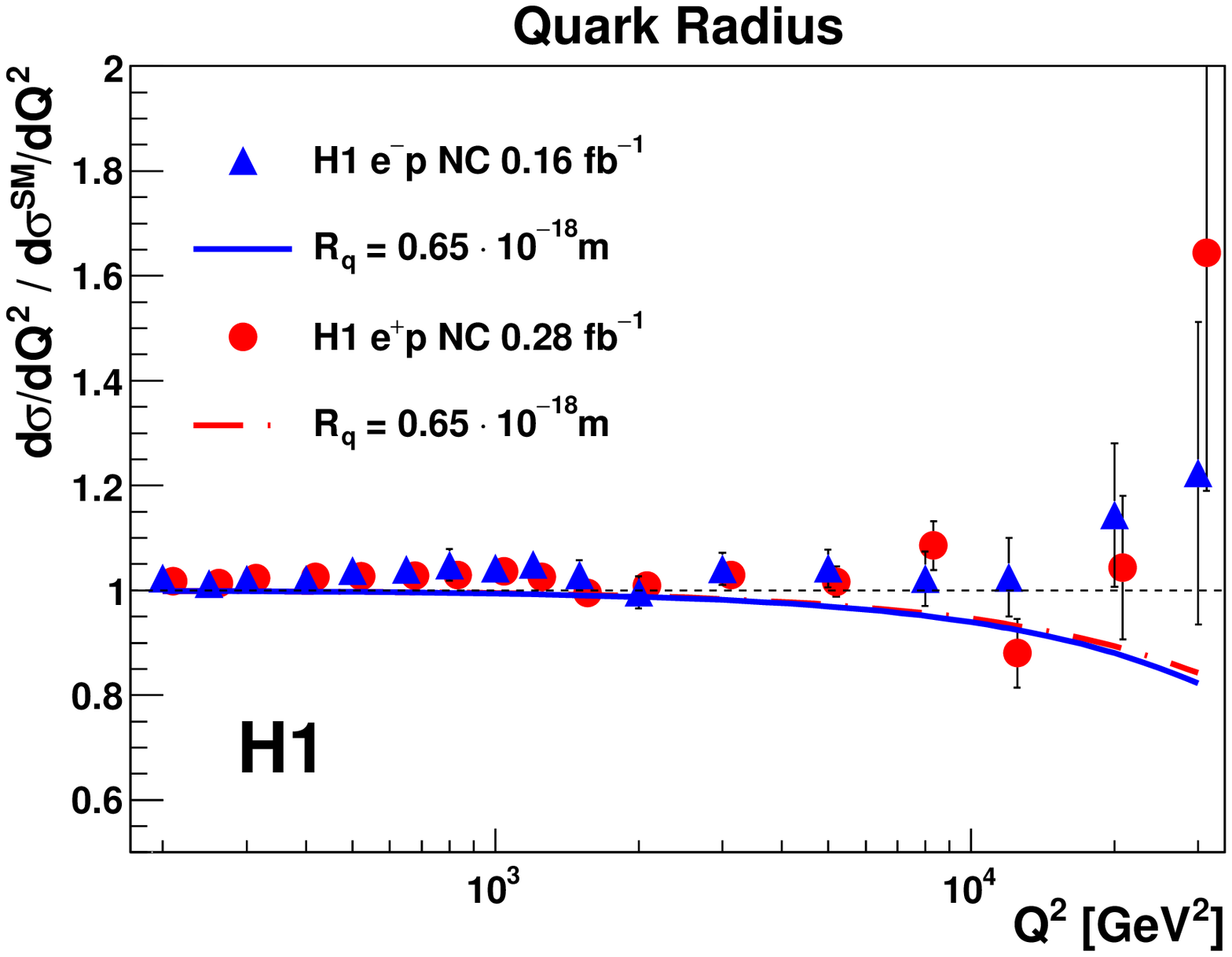}
\caption{The measured neutral current cross section ${\rm d}\sigma/{\rm d}Q^2$
  	normalised to the Standard Model expectation. H1 $e^{\pm}p$ scattering
	data are compared with curves corresponding to $95\%$~CL exclusion
	limits obtained from the full H1 data on the quark radius, $R_q$
	assuming point-like leptons. The error bars represent the statistical
	and uncorrelated systematic errors added in quadrature.}
\label{fig:dsdq2normrad}
\end{figure}


\begin{thebibliography}{99}



% H1 CI paper
\bibitem{h1ci2003} C.~Adloff {\it et al.}  [H1 Collaboration],
%        ``Searches and electroweak measurements at HERA,''
        Phys.\ Lett.\ B {\bf 568} (2003) 35 [hep-ex/0305015].
        %%CITATION = HEP-EX 0003002;%%

\bibitem{h1ci2000} C.~Adloff {\it et al.}  [H1 Collaboration],
%        ``Search for compositeness, leptoquarks and large extra dimensions in e q  contact interactions at HERA,''
        Phys.\ Lett.\ B {\bf 479} (2000) 358 [hep-ex/0003002].
        %%CITATION = HEP-EX 0003002;%%

\bibitem{zeusci2004}
  S.~Chekanov {\it et al.}  [ZEUS Collaboration],
%  ``Search for contact interactions, large extra dimensions and finite  quark
%  radius in e p collisions at HERA,''
  Phys.\ Lett.\  B {\bf 591} (2004) 23
  [hep-ex/0401009].
  %%CITATION = PHLTA,B591,23;%%

\bibitem{zeusci2000} J.~Breitweg {\it et al.}  [ZEUS Collaboration],
%        ``Search for contact interactions in deep inelastic e+ p $\to$ e+ X  scattering at HERA,''
        Eur.\ Phys.\ J.\ C {\bf 14} (2000) 239 [hep-ex/9905039].
        %%CITATION = HEP-EX 9905039;%%       

\bibitem{LEP} K.~Cheung, Phys.\ Lett.\  B {\bf 517} (2001) 167
  [hep-ph/0106251].


%%%



\bibitem{h1lfv}
  F.~D.~Aaron {\it et al.}  [H1 Collaboration],
%  ``Search for Lepton Flavour Violation at HERA,''
  accepted by Phys.\ Lett. B~[arXiv:1103.4938].

\bibitem{h1rpv}
 F.~D.~Aaron {\it et al.}  [H1~Collaboration],
%  ``Search for Squarks in R-parity Violating Supersymmetry in ep Collisions at
%  HERA,''
  Eur.\ Phys.\ J.\  C {\bf 71} (2011) 1572
  [arXiv:1011.6359].
% [hep-ex]].

\bibitem{atlaslq} G.~Aad {\it et al.}  [ATLAS Collaboration],
%  ``Search for pair production of first or second generation leptoquarks in
%  proton-proton collisions at $\sqrt{s}=7$ TeV using the ATLAS detector at the
%  LHC'', 
  submitted to Phys.\ Rev.\ D [arXiv:1104.4481].
% [hep-ex]].

\bibitem{cmslq} V.~Khachatryan {\it et al.} [CMS Collaboration],
%  ``Search for Pair Production of First-Generation Scalar Leptoquarks in pp
%  Collisions at $\sqrt{s}=7$'', 
  submitted to Phys.\ Rev.\ D [arXiv:1012.4031].
% [hep-ex]].

\bibitem{d0lq} V. Abazov {\it et al.} [D{\O} Collaboration]
Phys. Lett. B {\bf 671} (2009) 224 [arXiv:0907.1048].
% [hep-ex]].

%\cite{Abazov:2008as}
\bibitem{d0led2008}
  V.~M.~Abazov {\it et al.}  [D{\O} Collaboration],
%  ``Search for Large extra spatial dimensions in the dielectron and diphoton
%  channels in p anti-p collisions at s**(1/2) = 1.96-TeV,''
  Phys.\ Rev.\ Lett.\  {\bf 102} (2009) 051601
  [arXiv:0809.2813].
% [hep-ex]].
  %%CITATION = PRLTA,102,051601;%%

%\cite{:2009mh}
\bibitem{d0led2009}
  V.~M.~Abazov {\it et al.}  [D{\O} Collaboration],
%  ``Measurement of dijet angular distributions at sqrt{s}=1.96TeV and searches
%  for quark compositeness and extra spatial dimensions,''
  Phys.\ Rev.\ Lett.\  {\bf 103} (2009) 191803
  [arXiv:0906.4819].
% [hep-ex]].
  %%CITATION = PRLTA,103,191803;%%



%\cite{Chatrchyan:2011jx}
\bibitem{cmsled}
  S.~Chatrchyan {\it et al.}  [CMS Collaboration],
%  ``Search for Large Extra Dimensions in the Diphoton Final State at the Large
%  Hadron Collider,''
  JHEP {\bf 1105} (2011) 085
  [arXiv:1103.4279].
% [hep-ex]].
  %%CITATION = JHEPA,1105,085;%%






% CI phenomenology
\bibitem{elpr} E.~Eichten, K.~D.~Lane and M.~E.~Peskin,
%        ``New Tests For Quark And Lepton Substructure,''
        Phys.\ Rev.\ Lett.\  {\bf 50} (1983) 811; \\
        %%CITATION = PRLTA,50,811;%%
        R.~R\"uckl,
%        ``Effects Of Compositeness In Deep Inelastic Scattering,''
        Phys.\ Lett.\ B {\bf 129} (1983) 363; \\
        %%CITATION = PHLTA,B129,363;%%             
        R.~R\"uckl,
%        ``Probing Lepton And Quark Substructure In Polarized E-+ N Scattering,''
        Nucl.\ Phys.\ B {\bf 234} (1984) 91.
        %%CITATION = NUPHA,B234,91;%%

\bibitem{haberl} P.~Haberl, F.~Schrempp and H.-U.~Martyn,
        Proc. Workshop {\rm `Physics at HERA'}, 
        eds. W.~Buchm\"uller and G.~Ingelman,
        DESY, Hamburg (1991), vol. 2, p. 1133.

% LQ phenomenology



\bibitem{brw} W.~Buchm\"uller, R.~R\"uckl and D.~Wyler,
        Phys.\ Lett.\ B {\bf 191} (1987) 422 
        %erratum Phys.\ Lett.\ B {\bf 448} (1999) 320.
        [Erratum-ibid.\ B~{\bf 448} (1999) 320].
        %%CITATION = PHLTA,B191,442;%%

\bibitem{kalinowski} J.~Kalinowski, % {\it et al.},
        R.~R\"uckl, H.~Spiesberger and P.~M.~Zerwas,
%        ``Leptoquark/squark interpretation of HERA events: Virtual effects in e+  e- annihilation to hadrons,''
        Z.\ Phys.\ C {\bf 74} (1997) 595  [hep-ph/9703288].
        %%CITATION = HEP-PH 9703288;%%

\bibitem{rpv} J.~Butterworth and H.~Dreiner,
%        ``R-Parity Violation At Hera,''
        Nucl.\ Phys.\ B {\bf 397} (1993) 3 [hep-ph/9211204].
        %%CITATION = HEP-PH 9211204;%%

% quantum gravity
\bibitem{add} N.~Arkani-Hamed, S.~Dimopoulos and G.~R.~Dvali,
%        ``The hierarchy problem and new dimensions at a millimeter,''
        Phys.\ Lett.\ B {\bf 429} (1998) 263 [hep-ph/9803315]; \\
        %%CITATION = HEP-PH 9803315;%%
%        ``Phenomenology, astrophysics and cosmology of theories with  sub-millimeter dimensions and TeV scale quantum gravity,''
        N.~Arkani-Hamed, S.~Dimopoulos and G.~R.~Dvali,
	Phys.\ Rev.\ D {\bf 59} (1999) 086004 [hep-ph/9807344].
       %%CITATION = HEP-PH 9807344;%%

\bibitem{giudice} G.~F.~Giudice, R.~Rattazzi and J.~D.~Wells,
%        ``Quantum gravity and extra dimensions at high-energy colliders,''
        Nucl.\ Phys.\ B {\bf 544} (1999) 3
        [corrections in hep-ph/9811291 v2].
        %%CITATION = HEP-PH 9811291;%%                  

\bibitem{h1xsec} C.~Adloff {\it et al.}  [H1 Collaboration],
%        ``Measurement of neutral and charged current cross-sections in positron  proton collisions at large momentum transfer,''
        Eur.\ Phys.\ J.\ C {\bf 13} (2000) 609 [hep-ex/9908059].
        %%CITATION = HEP-EX 9908059;%%
\bibitem{h1xe-p} C.~Adloff {\it et al.}  [H1 Collaboration],
%        ``Measurement of neutral and charged current cross sections in electron  proton collisions at high $Q^2$,''
        Eur.\ Phys.\ J.\ C {\bf 19} (2001) 269 [hep-ex/0012052].
%        %%CITATION = HEP-EX 0012052;%%
\bibitem{h1xe+p} C.~Adloff {\it et al.}  [H1 Collaboration],
%         ``Measurement and QCD analysis of neutral and charged current cross  sections at HERA,''
         Eur.\ Phys.\ J.\ C {\bf 30} (2003) 1 [hep-ex/0304003].
%         %%CITATION = HEP-EX 0304003;%%




\bibitem{DGLAP}
V. N. Gribov and L. N. Lipatov, Sov. J. Nucl. Phys. \bf{15} \rm (1972) 438; \newline
V. N. Gribov and L. N. Lipatov, Sov. J. Nucl. Phys. \bf{15} \rm (1972) 675; \newline
L. N. Lipatov, Sov. J. Nucl. Phys. \bf{20} \rm (1975) 94; \newline
Y. L. Dokshitzer, Sov. Phys. JETP \bf{46} \rm (1977) 641; \newline
G. Altarelli and G. Parisi, Nucl. Phys. B \bf{126} \rm (1977) 298.

\bibitem{cteq6} J.~Pumplin {\it et al.}, % [CTEQ Collaboration],
       %J.~Pumplin, D.~R.~Stump, J.~Huston, H.~L.~Lai, P.~Nadolsky and W.~K.~Tung,
%       ``New generation of parton distributions with uncertainties from global  QCD analysis,''
       JHEP {\bf 0207} (2002) 012  [hep-ph/0201195].
       %%CITATION = HEP-PH 0201195;%%




\bibitem{h1lowx}
%\cite{Collaboration:2009bp}
 F.~D.~Aaron {\it et al.} [H1~Collaboration],
%  ``Measurement of the Inclusive ep Scattering Cross Section at Low $Q^2$ and x
%  at HERA,''
  Eur.\ Phys.\ J.\  C {\bf 63} (2009) 625
  [arXiv:0904.0929].
% [hep-ex]].
  %%CITATION = EPHJA,C63,625;%%

\bibitem{theses} 
R.~Pla\v{c}akyt\.{e}, ``First Measurement of Charged Current Cross Sections with
Longitudinally Polarized Positions at HERA'', PhD thesis,
University Munich (2006),
DESY-THESIS-2006-006
(available at http://www-h1.desy.de/publications/theses list.html); \newline
%
T.H.~Tran, ``Precision measurements of the charged and neutral current processes at high Q2 at HERA with polarized electron beam'', PhD thesis, University Paris-Sud (2010),  LAL-10-28.



\end{thebibliography}
\end{document}